# Evaluation of Zadoff-Chu, Kasami and Chirp based encoding schemes for Acoustic Local Positioning Systems

Santiago Murano, Carmen Pérez-Rubio, David Gualda, Fernando J. Álvarez,
Teodoro Aguilera and Carlos De Marziani

*Abstract*—The task of determining the physical coordinates of a target in indoor environments is still a key factor for many applications including people and robot navigation, user tracking, location-based advertising, augmented reality, gaming or emergency response. Among the different possibilities for indoor positioning, Acoustic Local Positioning Systems (ALPS) have the potential for centimeter level positioning accuracy with coverage distances up to tens of meters. In addition, acoustic transducers are small, low cost and reliable thanks to the room constrained propagation of these mechanical waves. Waveform design (coding and modulation) is usually incorporated into these systems to facilitate the detection of the transmitted signals at the receiver. The aperiodic correlation properties of the emitted signals have a large impact on how the ALPS cope with common impairment factors such as multipath propagation, multiple access interference, Doppler shifting, near-far effect or ambient noise. This work analyzes three of the most promising families of codes found in the literature for ALPS: Kasami codes, Zadoff-Chu and Orthogonal Chirp signals. The performance of these codes is evaluated in terms of time of arrival accuracy and characterized by means of model simulation under realistic conditions and in real-world scenarios. The results derived from this study can be of interest for other applications based on spreading sequences, such as underwater acoustic systems, ultrasonic imaging or even CDMA communications systems.

## I. Introduction

ACOUSTIC local positioning systems (ALPS) are based on the emission of sound or ultrasound signals that are usually transmitted in a bursting manner, and whose Times of Arrival (TOA) must be estimated with high accuracy, regardless the final positioning algorithm being based on absolute measurements of TOA or on Differences in Times of Arrival (DTOA) [1]. To enhance the robustness and accuracy in TOA estimation many efforts have been done during the last decades in the design of spreading sequences with good aperiodic properties, able to cope with typical sound signal impairments such as strong multipath, near-far effect, acoustic noise, bursting transmissions, asynchronous detection or Doppler shift arising from a moving emitter or receiver. Furthermore, to provide a high update rate, most systems require simultaneous emissions from different transducers, so the sequences must present low cross-correlation values among them to be uniquely identified at the receiver.

The study on spreading codes for ALPS or other Code Division Multiple Access (CDMA) based applications is a traditional research topic. [2] presents a revision of some of the most interesting aperiodic spreading sequences used for active sensing. From the first works by Hazas and Ward with Gold codes [3], many other have arisen. Some of them are also based on unitary spreading pseudo-random codes, such as m-sequences or Kasami codes [4], [5]. Despite they are designed for periodic correlation, the small set of Kasami codes seem to offer acceptable aperiodic properties and can be often found in ALPS proposals [1], [6], [7]. Other authors rely on multi-sequence codes that work on a multiple-sequence-per-user basis, as is the case of Golay pairs and Complementary Set of Sequences (CSS) [8], [9]. The sum of aperiodic correlation functions of the sequences of each set is ideal, with no auto-correlation or cross-correlation sidelobes. In practice, the set of sequences that encodes each transmitter is usually interleaved or concatenated to form a large unique sequence that is then transmitted through the transducer by means of a Binary Phase Shift Keying (BPSK) modulation. Alternatively, a more complex modulation can be used to simultaneously emit all sequences of the complementary set. In both cases, the inherent ideal properties of the codes are lost, so the sum of correlation functions is not interference-free [10]. Also, [11] demonstrates that CSS, and by extent generalized orthogonal codes derived from CSS (Inter-Group Complementary Codes [12], Loosely Synchronous [13], Generalized Pairwise Complementary codes [14]), are not a good choice when dealing with the Doppler effect, since the correlation sidelobes increase significantly with the receiver velocity. In case of multilevel CSS, new problems appear due to the requirement of high power amplifiers [2].

Recent works on ultrasonic ranging have relied on polyphase sequences. Specifically, Zadoff-Chu (ZC) codes are a good choice because of their length flexibility, good performance in periodic systems and tolerance to Doppler effect [15]. ZC sequences are polyphase complex valued sequences that have constant amplitude and zero auto-correlation side-

S. Murano, C. Pérez-Rubio and D. Gualda are with the Electronic Department at the University of Alcalá, Spain. (e-mail: santiago.murano@edu.uah.es, mcarmen.perezr@uah.es)

F. J. Álvarez and T. Aguilera are with the Sensory Systems Research Group, at the University of Extremadura, Spain. (e-mail: fafranco@unex.es, teoaguibe@unex.es)

S. Murano and C. De Marziani are with the Electronic Department at the National University of Patagonia San Juan Bosco, Argentina.

lobes when they are emitted periodically [16], [17]. They are currently used in the Long Term Evolution (LTE) air interface for radio communication systems as the primary synchronization signal. Also, they have been successfully employed in underwater acoustic systems [18], [19]. In a previous work, [20], the authors have compared the performance of ZC codes in aperiodic emission, evaluating four different modulation schemes: Quaternary Phase Shift Keying (QPSK), Fast Frequency Hopping Spread Spectrum (Fast-FHSS), Slow-FHSS and Orthogonal Frequency Division Multiplexing (OFDM). Results showed that ZC codes with QPSK modulation, non coherently detected by matched filtering, offers better accuracy than the other modulation schemes. In practice, modulation is required to adjust the codes to be transmitted to the available bandwidth of the channel which is constrained by the transducers.

Other researchers have also focused on the search of the appropriate modulation schemes for ALPS. In [6] is analyzed the performance of BPSK, Fast-FHSS and Slow-FHSS in the transmission of Kasami codes in an ALPS that uses both TOA and Angle of Arrival (AoA). BPSK modulation allows for an efficient use of the bandwidth, but it causes large sidelobes in the surroundings of the main correlation peak so taking the time of the maximum value for the TOA estimation can entail a small error. On the other hand, FHSS-based ALPS seem to offer better accuracy and noise performance since the carrier is hopping in frequency and only noise and multipath in the same band affects it. Nevertheless, the modulation is less spectrally efficient, the effective bandwidth for every hop is significantly reduced when the number of simultaneous emissions increase, the electronic is more complex and require more restrictive synchronization issues.

OFDM is another modulation approach that is mainly used in underwater acoustic communication systems [19], [21], and has also been extended to ultrasonic ALPS applications [22]. However, OFDM usually suffers from high peak-to-average power ratio and the Doppler effect can break the subcarriers orthogonality. Very recently, Orthogonal Chirp Division Multiplexing (OCDM) has been introduced in ultrasonic positioning [23], [24]. OCDM consists of using orthogonal chirp waveforms currently used in radar [25] and data communications [26] to modulate the transmitted signals with significantly low interference, while maintaining the properties of classic chirp signals.

Motivated by the call for innovation in ALPS to provide services based on position for mobile robots, people or electronics devices, this work intends to evaluate the most promising encoding-modulation schemes found in the literature in real scenarios, considering typical impairing factors. This work extends our preliminary results in [20]. Based on these previous results, we have discarded FHSS due to their worst tolerance to Doppler effect and, in addition to QPSK and OFDM, we have included here other relevant alternatives: Kasami codes that have been already demonstrated to be a good choice for general-purpose ALPS, and a novelty OCDM modulating scheme for ZC codes. Also, the experimental tests have been extended including experiments regarding the Doppler effect and with an ultrasonic ALPS operating in a wider area. A systematic analysis has been carried out which includes real scenario common hindering effects such as white noise conditions, near-far, multipath, multiple access and Doppler shift. These impairment factors have been analyzed both by simulation and experimental testing in terms of TOA accuracy estimation. Also, experimental results are shown in an ultrasonic ALPS where the beacons work as emitters and a non limited number of receivers can compute their position, thus maintaining privacy. To the best of our knowledge this is the first paper to report on such a comparison and could be very useful to prevent from needing complex inter-symbol-interference and multiple-access-interference compensation algorithms.

The rest of the paper is organized as follows. Section II reviews the main properties of the encoding and modulation schemes under study. Section III is focused on the specific parameters in the waveform design of the different encoding-modulation schemes for a fair comparison between them. Section IV reviews the most common effects in ALPS, and Section V presents the formulation used in the simulation of the effects and the results obtained for every encoding scheme. Section VI describes the experimental performance in two situations: an ultrasonic ALPS with five emitting beacons and a single emitter-receiver scheme affected by Doppler effect. Finally, these results are discussed in Section VII, where the main conclusions of the work are also drawn.

## II. SIGNAL DESIGN

This section introduces the encoding and modulation schemes under analysis.

### A. Encoding schemes

*1) Binary codes:* Kasami sequences [5] belong to the family of Pseudo Noise (PN) sequences, they are composed by $K_{kas} = 2^{N_{kas}/2}$ binary sequences, with length $L_{kas} = 2^{N_{kas}} - 1$, where $N_{kas}$ is an even natural number. A new Kasami sequence $KAS_k[n]$ can be obtained from a maximal sequence and the decimated and concatenated version of this sequence by performing the modulo-2 sum of the former with any delayed version of the latter, i.e.,

$$KAS_k[n] = m_1[n] \oplus D^l m_2[n] \quad \text{with } l < L_{kas} \quad (1)$$

where $m_1$ is a maximal sequence of length $L_{kas}$, $m_2$ is obtained by decimating $m_1$ with a decimation factor of $q = 2^{N_{kas}/2} + 1$ and then concatenating the result $q$ times, $\oplus$ represents the modulo-2 sum, and $D^l m_2$ is the sequence obtained by cyclically shifting $l$ positions the $m_2$ sequence.

*2) Polyphase codes:* Zadoff-Chu (ZC) sequences, are polyphasic and unitary complex valued codes. According to [16] [17], a ZC sequence $Z_{r_k}$ of length $L_{ZC}$, is defined as

$$Z_{r_k}[l] = \begin{cases} \exp\left(j\pi \frac{r_k l^2}{L_{ZC}}\right), & \text{if } L_{ZC} \text{ is even} \\ \exp\left(j\pi \frac{r_k l(l+1)}{L_{ZC}}\right), & \text{if } L_{ZC} \text{ is odd} \end{cases} \quad (2)$$

where $r_k = 1, 2, \ldots, L_{ZC} - 1$ is an integer and co-prime to $L_{ZC}$ that represents the generation root for each of the $K_{ZC} = L_{ZC} - 1$ available sequences. The periodic Auto-Correlation (AC) function of a ZC sequence is defined as

$$R_{Z_{r_k}}(t) = \sum_{l=0}^{L_{ZC}-1} Z_{r_k}(l) \tilde{Z}_{r_k}^*(l+t) \quad (3)$$

where $\tilde{Z}_{r_k}^*(l)$ denotes the periodically repeated conjugated sequence $Z_{r_k}(l)$. In [17] it was shown that the periodic AC function has zero sidelobes, as indicated in (4).

$$R_{Z_{r_k}}(t) = \begin{cases} L_{ZC}, & t \bmod L_{ZC} = 0, \\ 0, & t \bmod L_{ZC} \neq 0, \end{cases} \quad (4)$$

On the other hand, the aperiodic AC function can be defined as [27]:

$$C_{Z_{r_k}}(t) = \begin{cases} \sum_{l=0}^{L_{ZC}-1-t} Z_{r_k}(l) Z_{r_k}^*(l+t) & \text{if } t \geq 0, \\ \sum_{l=-t}^{L_{ZC}-1} Z_{r_k}(l) Z_{r_k}^*(l+t) & \text{if } t < 0. \end{cases} \quad (5)$$

where $Z_{r_k}^*()$ denotes the complex conjugate of $Z_{r_k}()$. The AC function at $t = 0$ is equal to the length $L_{ZC}$ of the sequence. The aperiodic correlation of ZC sequences is not ideal, but achieves low correlation sidelobes. According to [27], the maximum Sidelobe to Mainlobe Ratio (SMR) in aperiodic emission is approximately $1/(2.085\sqrt{L_{ZC}}-0.0736)$ when $L_{ZC} < 45$. The SMR can be defined as:

$$\text{SMR} = \frac{\max_{1 \leq t \leq L_{ZC}-1}\{|C_{Z_{r_k}}(t)|\}}{L_{ZC}} \quad (6)$$

*3) Chirp sequences:* In a linear chirp, the instantaneous frequency $f(t)$ varies linearly with time $t$, considering a bandwidth $B = f_h - f_l$, where $f_l$ and $f_h$ are the initial and final frequencies respectively, during a time period $T_c$. The time-domain sinusoidal linear chirp can be defined as:

$$CH(t) = \sin\left(2\pi\left(f_{st}t + \frac{1}{2}\mu t^2\right) + \varphi_l\right) \quad 0 \leq t \leq T_c \quad (7)$$

where $f_{st}$ is the start frequency, $\mu = B/T_c$ is the chirp rate, and $\varphi$ the initial phase.

Traditional chirp codes suffer from the multiple-access problem [28]. To support multiple access a new modulating scheme called OCDM has been recently introduced in the field of optical fiber communications [29], and latterly used in ultrasonic systems [23], [24]. Here, we will consider the proposal in [23] which consists of the simultaneous emission of quasi-orthogonal chirp waveforms that can be expressed as:

$$CH_k(t) = CH_p(t) + CH_q(t) =$$
$$\text{rect}\left(\frac{t}{T_{c_p}}\right) \exp\left(j 2\pi\left(f_{st_{k_p}}t + \frac{1}{2}\mu_{k_p}t^2\right)\right) + \text{rect}\left(\frac{t-T_{c_p}}{T_{c_q}}\right)$$
$$\times \exp\left(j 2\pi\left(f_{st_{k_q}}(t-T_{c_p}) + \frac{1}{2}\mu_{k_q}(t-T_{c_p})^2\right)\right) \quad (8)$$

where $k = 1, \ldots, K_{CH}$ is related to the $k$-th waveform generated, $\text{rect}[\ ]$ is a rectangular window function. The waveform is obtained by concatenating two partial chirp signals $CH_p(t)$ and $CH_q(t)$, each with their respective starting frequency $f_{st_{k_p}}$ or $f_{st_{k_q}}$, frequency rate $\mu_{k_p}$ or $\mu_{k_q}$, and chirp duration $T_{c_p}$ or $T_{c_q}$, respectively. Note that $T_c = T_{c_p} + T_{c_q}$.

### B. Modulation schemes

The encoded signals must be modulated before their emission to fit the frequency response of the ultrasonic transducer. This section introduces the different schemes utilized, based on phase modulation such as BPSK and QPSK, OFDM, and chirp based modulation (OCDM).

*1) Phase Shift Keying:* An easy way to transmit binary Kasami codes is by means of a BPSK modulation. This scheme has been widely employed in different location or range systems [1], [3], [11]. One or more carrier cycles are used to modulate each bit in the code $KAS_k[n]$, so the final phase 0 or $\pi$ will depend on the bit value as:

$$sk_{kas}[n] = \sum_{l=0}^{L_{kas}-1} KAS_k[l] \cdot \sin(2\pi f_c n) \quad (9)$$

$$0 \leq n/f_s \leq O_f O_c/f_s$$

where $sk_{kas}[n]$ is the modulated Kasami code; $O_c$ is the number of carrier cycles; and $O_f = \frac{f_s}{f_c}$ represents the oversampling factor (ratio between the sampling frequency $f_s$, and carrier frequency $f_c$).

On the other hand, one of the alternatives analyzed for the transmission of the complex values of ZC sequences is based on a Quadrature Phase Shift Keying (QPSK) scheme. Both the real and imaginary parts are phase modulated by using orthogonal carriers, such as cosine and a sine wave signals. Then, the emitted signal will be the sum of both, as can be seen in expression (10):

$$C_{f_c}[n] = \cos(2\pi f_c n) - j \cdot \sin(2\pi f_c n)$$
$$0 \leq n/f_s \leq O_f O_c/f_s$$

$$sk_{QPSK}[n] = \sum_{l=0}^{L_{ZC}-1} \text{Re}(Z_{r_k}[l]) \cdot \text{Re}(C_{f_c}[n])$$
$$- \text{Im}(Z_{r_k}[l]) \cdot \text{Im}(C_{f_c}[n]) \quad (10)$$

where $C_{f_c}[n]$ is the complex carrier, centered at a frequency $f_c$, with a symbol duration that depends on the oversampling factor $O_f$ and the number of cycles per symbol $O_c$.

*2) OFDM:* Orthogonal Frequency Division Multiplexing (OFDM) modulation is a multicarrier medium access technique that distributes data over $N_{OFDM}$ equally spaced frequency channels, in the bandwidth from 0 to $\frac{f_s}{2}$ Hz, where $f_s$ is the sampling frequency. The use of orthogonal sub-carriers allows overlapping, what improves the bandwidth usage and increases the spectral efficiency compared to other multi-carrier modulation schemes [30]. This modulation scheme can be efficiently implemented by using the Inverse Fast Fourier Transform (IFFT) at the emitter, and the FFT at the receiver. The Discrete Multitone Modulation (DMT) [31] is the same as OFDM but it uses instead an IFFT with the double number of samples, i.e. $N_{DMT} = 2N_{OFDM}$. It also takes advantage

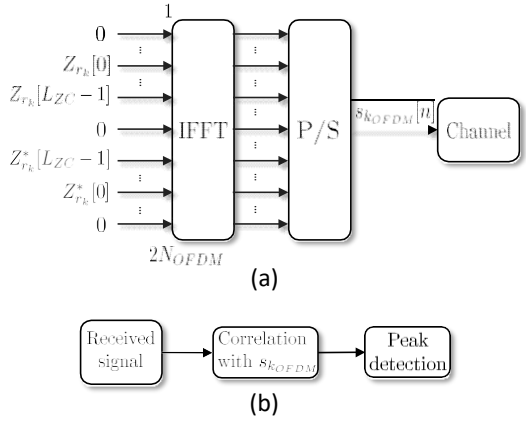

Fig. 1: OFDM schemes, using a ZC sequence for the $i$-th emitter: (a) emitter and (b) receiver.

of the Hermitian symmetry property to output a real signal. This modulation can be expressed as:

$$s_{k_{OFDM}}[n] = \sum_{l=0}^{2N_{OFDM}-1} S_k[l] \cdot e^{j2\pi nl/2N_{OFDM}};$$
$$n = 0, \ldots, 2N_{OFDM} - 1 \quad (11)$$

where $S_k[2N_{OFDM} - l] = S_k[l]^*$, being $S_k[l]^*$ the complex conjugate of $S_k[l]$, that indicates the symbol from emitter $k$, transmitted in the sub-band $l$. Taking into account the DMT bandwidth split, to allocate the Zadoff-Chu code in the appropriate band defined by the transducer response, each code bit has to be fed into the channels of the IFFT that correspond to the desired output frequencies, and the rest of the channel inputs should be filled with zeros. The number of zeros to be added at the lower and higher part are $Z_L$ and $Z_H$ respectively, and they are calculated as:

$$Z_H = \frac{L_{ZC}(f_s - 2f_h)}{2(f_h - f_l)}, \quad Z_L = \frac{2f_l(L_{ZC} + Z_H)}{f_s - 2f_l} \quad (12)$$

This number of zeros depends on the sampling rate $f_s$, the available total bandwidth $B = f_h - f_l$, and the sequence length $L_{ZC}$.

The input data $S_k[l]$ in (11), before applying the Hermitian symmetry, is a vector of length $N_{OFDM}$:

$$S_k[l] = [0, zeros(1:Z_L), Z_{r_k}[0], \ldots,$$
$$, Z_{r_k}[L_{ZC} - 1], zeros(1:Z_H)] \quad (13)$$

Fig. 1 shows the emitter and receiver schemes. At the receiver, the incoming signal is directly correlated with the modulated one $s_{k_{OFDM}}[n]$.

*3) Chirp (OCDM):* As a novel approach, a set of symbols based on the generation of orthogonal and complex chirp sequences (see Section II-A3), are used as symbols to modulate a set of ZC sequences, in a similar way than the QPSK case explained before. The resulting signal can be written as:

$$s_{k_{CH}}[n] = \sum_{l=0}^{L_{ZC}-1} \text{Re}(Z_{r_k}[l]) \cdot \text{Re}(CH_k[n])$$
$$- \text{Im}(Z_{r_k}[l]) \cdot \text{Im}(CH_k[n]) \quad (14)$$

### III. CONFIGURED PARAMETERS

For a fair comparison between the different encoded-modulated schemes under evaluation, all of them should have similar emitted energy and time duration. The signals have been designed considering the beacon unit used later in the real-world tests, which is further described in Section VI-A. It consists of five ultrasonic transducers ($K_b = 5$) that are hardware synchronized and simultaneously emit the generated sequences. The transducers have an approximate bandwidth of 8 kHz centered at 41.67 kHz. The sampling frequency for the emission is $f_s$ = 500 kHz. Considering that, in each case the family size should be greater than five to unequivocally identify every transducer. The encoding and modulation schemes are generated as follows:

*1) Kasami - BPSK (Kas-BPSK):* A set of five Kasami sequences of length $L_{kas}$ = 255 BPSK modulated with an oversampling factor $O_f$ = 12 and with a number of cycles per symbol $O_c$ = 4.

*2) Zadoff-Chu - QPSK (ZC-QPSK):* A set of five ZC sequences were generated with a prime length $L_{ZC}$ = 257, and generation roots $r_k$ = [1, 256, 129, 128, 86] respectively for each obtained sequence. They were QPSK modulated with an oversampling factor $O_f$ = 12 and a number of cycles per symbol $O_c$ = 4.

*3) Zadoff-Chu - OFDM (ZC-OFDM):* A set of five ZC sequences were generated, with a prime length $L_{ZC}$ = 191, and generation roots $r_k$ = [1, 2, 190, 189, 3]. Then, they are OFDM modulated, adding $Z_H$ = 883 and $Z_L$ = 4894 zeros before applying the IFFT. Note that since the length of the ZC sequences is different than the one of the QPSK modulation scheme, the roots have changed accordingly to obtain ZC sequences with low Cross-Correlation (CC) among them. For low CC, the roots should be co-primes and primes with $L_{ZC}$.

*4) Zadoff-Chu - OCDM (ZC-Chirp):* A set of five ZC sequences were generated, with a prime length $L_{ZC}$ = 67, and generation roots $r_k$ = [1, 66, 34, 33, 22]. To modulate them, five orthogonal chirp pulses were obtained as indicated in (8) and by considering $f_l$ = 37kHz, $f_h$ = 45kHz, $T_c$ = 0.36ms where $T_f = T_g = \frac{T_c}{2}$, $\mu = B/T_c$. Table I shows the starting frequencies and chirp rates used.

Fig. 2 depicts the resultant time-frequency diagram for the proposed OCDM signals, that are used as modulation symbols for the transmission of the 67 bit ZC codes.

It is important to remark that not only the signal duration, but also the energy emitted to the channel should be similar for a fair comparison. Thus, the energy has been equalized. First, the signals are normalized in amplitude and then filtered with a Finite Impulse Response (FIR) model of the ultrasonic transducer. After that, the energy of the filtered signals is calculated with Eq. (15), where $N$ is the modulated signal length. Then, an attenuation factor $0 \leq AF_k \leq 1$ is computed for every $0 \leq k \leq K_b$ signal to be transmitted by considering the one with less energy as a reference. Lastly, this attenuation factor $AF_k$ is applied to each signal to be transmitted according to (17).

TABLE I: Starting frequencies and chirp rates for OCDM signal $s_{k_{CH}}$ generation.

|  | $f_{st_{k_p}}$ | $f_{st_{k_q}}$ | $\mu_{k_p}$ | $\mu_{k_q}$ |
|---|---|---|---|---|
| $s1_{CH}$ | $f_l$ | $f_l + \frac{B}{4}$ | $\frac{\mu}{2}$ | $\frac{3\mu}{2}$ |
| $s2_{CH}$ | $f_h$ | $f_h - \frac{B}{4}$ | $-\frac{\mu}{2}$ | $-\frac{3\mu}{2}$ |
| $s3_{CH}$ | $f_l$ | $f_l + \frac{3B}{4}$ | $\frac{3\mu}{2}$ | $\frac{\mu}{2}$ |
| $s4_{CH}$ | $f_h$ | $f_h - \frac{3B}{4}$ | $-\frac{3\mu}{2}$ | $-\frac{\mu}{2}$ |
| $s5_{CH}$ | $f_l$ | $f_l + \frac{B}{2}$ | $\mu$ | $\mu$ |

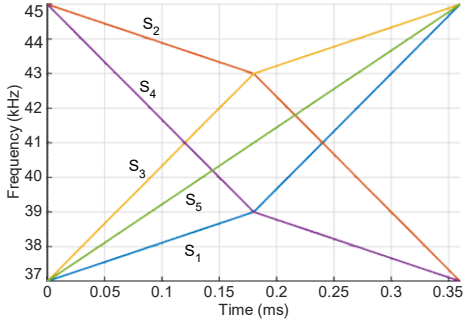

Fig. 2: Set of five linear chirp signals, generated with a time duration of $T_c = 0.36$ ms

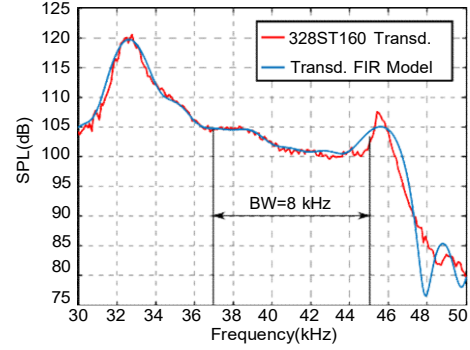

Fig. 3: 328ST160 transducer frequency spectrum, measured (red) and the FIR filter model (blue).

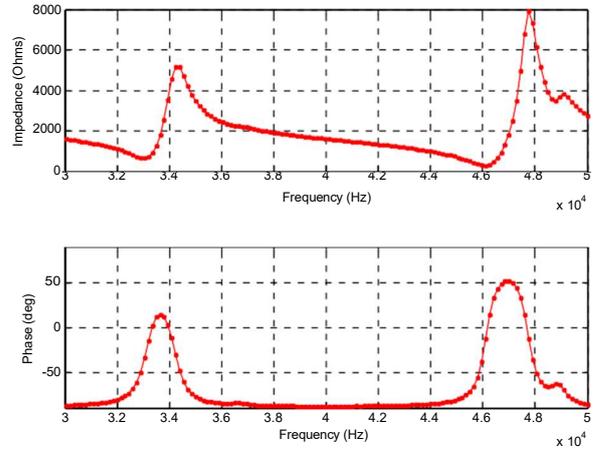

Fig. 4: 328ST160 transducer Impedance/Phase angle vs. frequency.

$$E_k = \frac{\sum_{n=0}^{N-1} \hat{s}_k[n]^2}{N} \quad (15)$$

$$AF_k = \sqrt{\frac{\min(E_{BPSK}, E_{QPSK}, E_{OFDM}, E_{Chirp})}{E_k}} \quad (16)$$

$$Tx_k[n] = s_k[n] \cdot AF_k \quad (17)$$

The main features of the resultant signals for a fair comparison are summarized in Table II. In the BPSK and QPSK modulation schemes, the frequency range was obtained considering the main frequency peak at $-3$ dB.

TABLE II: Signals under evaluation.

| Code / Modulation | Tx time (ms) | Sequence length | $AF_k$ | Freq. range (kHz) |
|---|---|---|---|---|
| Kas-BPSK | 24.48 | 255 | 0.58 | 37-46 |
| ZC-QPSK | 24.67 | 257 | 0.61 | 37-46 |
| ZC-OFDM | 23.87 | 191 | 1.00 | 37-45 |
| ZC-Chirp | 24.25 | 67 | 0.57 | 37-45 |

## IV. COMMON EFFECTS IN ALPS

This section provides the necessary background related to the common effects on air positioning systems. It is complemented with Section V that deals with the modelling of the effects for their simulation and shows the results obtained for each encoding and modulation scheme.

### A. Frequency/Phase response of the transducer

The emitters used for the experimental tests are based on the Prowave 328ST160 transducer [32], whose frequency spectrum is represented in Fig. 3 with a red line. For simulation purposes, a FIR filter was designed to simulate the transducer effect on the emitted signal. The obtained response is also represented in Fig. 3 with a blue line. In addition, the Impedance/Phase angle versus frequency representation can be seen in Fig. 4.

### B. Channel effects

The $K_b$ transmitters simultaneously emit their corresponding modulated signal $s_k(t)$, that will be affected by the transducer response $Th$ and the noise $\eta(t)$. At the reception, all signals coming from the transducers are received together, each one with its corresponding propagation delay $t_k$ and attenuation $A_k$, as indicated in Eq. (18).

$$r(t) = \sum_{k=1}^{K_b} A_k \cdot (Th * s_k)(t - t_k) + \eta(t) \quad (18)$$

where $*$ denotes the convolution operator.

The TOA $t_k$ depends on the speed of sound $c$ and the distance $d_k$ between the emitter and the receiver, as $t_k = d_k/c$.

The propagation speed of sound in air is influenced by the temperature as follows:

$$c = c_0 \cdot \sqrt{1 + \frac{T}{273.15}} \quad (19)$$

where $c_0 = 331.6$ m/s is the speed at zero Celsius and $T$ is the air temperature in Celsius.

### C. Effects conditioned by the receiver position

When the receiver is moving towards or away from a set of emitters, two effects should be considered. The first one is the near-far effect, which occurs when the received power difference between the incoming signals is so great that it is not possible the discrimination of the weakest signal. This effect can be modelled by modifying the attenuation factor $A_k$ of some emitters.

The second effect is the Doppler shift, which is related to the relative movement between the receiver and the emitters. This effect is easily modelled by assuming a virtual sampling frequency $f_s'$ for the emitted signal as follows [11]:

$$f_s' = f_s \left(1 - \frac{\vec{v}_r \cdot (\vec{r}_e - \vec{r}_r)}{c \, |\vec{r}_e - \vec{r}_r|}\right) \quad (20)$$

where $f_s$ is the actual sampling frequency, $c$ is the speed of sound in air, $\vec{r}_r$ is the receiver position vector, $\vec{r}_e$ is the emitter position vector, and $\vec{v}_r$ is the receiver velocity vector. From this frequency, the signal acquired by the receiver at the actual sampling frequency is obtained by an interpolation and decimation process.

Finally, an ideal propagation model only takes into account the direct-path received signal. Nevertheless, considering a realistic multipath environment, the received signal in Eq. (18) can be expanded to consider the reflections on multiple obstacles as:

$$r(t) = \sum_{k=1}^{K} \sum_{l=1}^{MP} A_{kl} \cdot (Th * s_k)(t - t_k - d_l) + \eta(t) \quad (21)$$

where $MP$ represents the number of replicas of the emitted signal $s_k$ with a propagation delay $d_l$ for the corresponding replica. It is important to mention that the cross-correlation peak associated with the direct-path is not always the one with the greatest amplitude, since multipath can produce a constructive sum of delayed signals resulting in a higher peak.

## V. SIMULATIONS

This section presents the simulation results for typical indoor scenarios. Every effect is separately analyzed, considering in all cases that the signals are emitted through Prowave 328ST160 transducers [32], as indicated in Section IV-A. In all simulations, the TOA is computed from the maximum peak of the aperiodic correlation between the affected signal and the emitted pattern $s_k(t)$. Note that more sophisticated peak detectors can improve the TOA estimations [33], but they also increase the complexity of the system. Our purpose is to identify the encoding schemes that better tackle all major adverse operational conditions in their code structure, thus avoiding the need of additional complex auxiliary algorithms for interference mitigation.

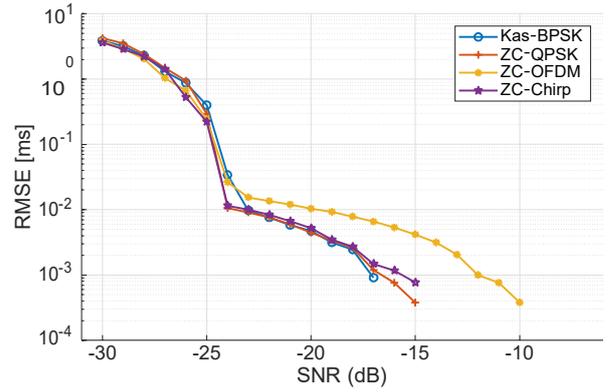

Fig. 5: RMSE in the peak detection as a function of the SNR variation for White Gaussian Noise simulation effect.

### A. White Gaussian noise

Environmental noise is always present, it could come from human or mechanical activities, and can affect TOA estimation depending on its energy and frequency. For instance, ultrasonic noise can appear in indoor environments due to the clacking of someone typing on a computer or the keyring jingle.

To test this behavior on the proposed schemes, Gaussian noise has been added to a single emitted signal (multi emission is not considered in this test) decreasing the SNR from $-5$ to $-35$ dB in steps of $1$ dB. Each scheme was run $5\,000$ iterations per SNR step. Fig. 5 represents the Root-Mean-Square Error (RMSE) in milliseconds in the detection of the main peak position. It is worth mentioning that null values are not represented because of the logarithmic scale.

### B. Near-Far effect

To simulate this effect, two signals ($s_1$ and $s_2$) have been generated. The first signal $s_1$ is received without attenuation, $A_1 = 1$, or displacement, $t_1 = 0$. On the contrary, the second signal is multiplied by a $0 \leq A_2 \leq 0.5$ attenuation factor, that is increased in steps of $0.01$, and it is received with a $t_2$ random delay between $0.4$ ms and $2$ ms. Each scheme was run for $5\,000$ iterations per $A_2$ step, considering in each iteration the aperiodic correlation between the affected signal and the emitted pattern $s_k(t)$. The received signal can be expressed as:

$$r(t) = (Th * s_1)(t) + A_2 \cdot (Th * s_2)(t - t_2) + \eta(t) \quad (22)$$

The tests were conducted in a SNR $= 20$ dB scenario. The RMSE in milliseconds in the detection of the correlation main peak position is shown in Fig. 6, as a function of the attenuation factor $A_2$. When $A_2 > 0.5$, that is, when the signal $s_2$ is received with more than half of its original power, all codes except OFDM result on null RMSE error. In case of OFDM, the CC between sequences lead to errors lower than $1.5\mu s$ when $A_2 > 0.5$.

### C. Multipath effect

According to Eq. (21), in indoor environments a receiver usually detects multiple delayed and attenuated replicas of a transmitted signal due to the specular reflections from walls and objects. To simulate this effect, one emitter $s_1$ has been considered and 10 000 iterations were performed per each encoding scheme. In every iteration, it is possible to find the direct path ($t_1 = 0$) and a set of $MP = 10$ random multipaths whose delays in milliseconds, $d_l$, and attenuation factors, $A_{1l}$, are obtained from the decreasing exponential function depicted in Fig. 7. Note that the delay is constrained between the range $d_l \in [0.4, 2]$ ms. Again, a SNR = 20 dB is considered.

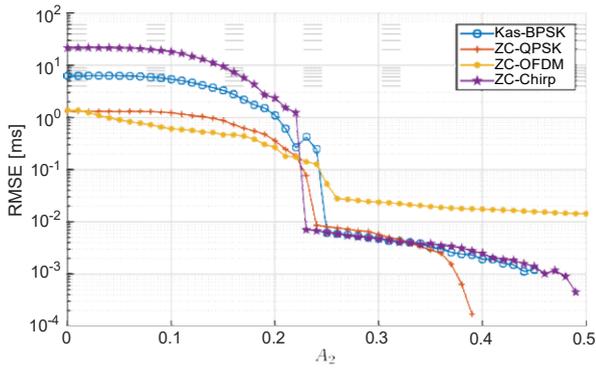

Fig. 6: RMSE in the peak detection considering the near-far effect, as a function of the attenuation factor $A_2$.

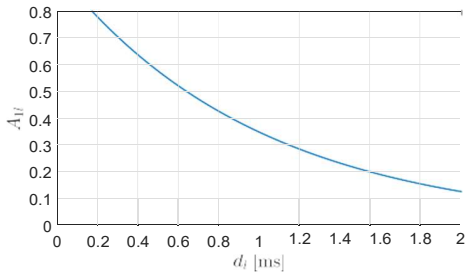

Fig. 7: Exponential attenuation factor $A_{1l}$ regarding to the random selected delay $d_l$.

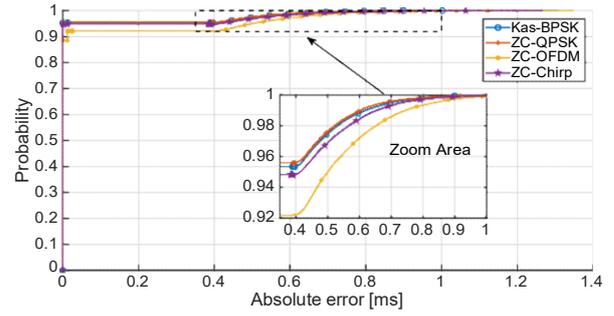

Fig. 8: Cumulative distribution function of the peak detection, for a signal affected by ten random multipaths.

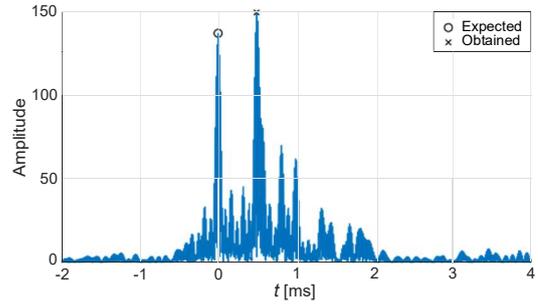

Fig. 9: Example of wrong correlation peak detection due to a constructive sum of several multipaths, for a BPSK modulated Kasami code.

$$r(t) = \sum_{l=1}^{10} A_{1l} \cdot (Th * s_1)(t - d_l) + \eta(t) \quad (23)$$

Fig. 8 shows the Cumulative Distribution Function (CDF) of the peak detection absolute error at the receiver for each encoding-scheme. Less than 10% of the results present an error on the peak detection. These errors are due to a constructive sum of several multipath as can be seen in the correlation example presented in Fig. 9.

### D. Doppler effect

The relative movement between the emitter and the receiver produces a signal expansion or compression, which degrades the received signal and produces in some cases the loss of the correlation main peak. For simulation purposes, Eq. (20) was considered to generate a virtual sampling frequency at the receiver.

Tests at different speeds were carried out considering only the transducer effect, and a SNR = 20 dB, with neither multipath nor cross-correlation effects. The speed is increased from 0 to 4 m/s at steps of 0.01 m/s, performing 5 000 iterations per step.

The RMSE related to the peak detection error in milliseconds versus the relative speed is shown in Fig. 10.

The SMR can be used as a measurement of the difficulty to detect the correlation peak in order to estimate the TOA. A guard factor $N_G$ around the mainlobe can be used to avoid confusing the maximum sidelobe with the lobes corresponding to the modulation effect. Thus, Eq. (6) can be written as:

$$\text{SMR} = \frac{\overline{\max(C_{r,s_k}[n]); \{\forall n \notin [-N_G, N_G]\}}}{\max(C_{r,s_k}[n]); \{\forall n\}} \quad (24)$$

where $C_{r,s_k}$ represents the aperiodic correlation function be-tween the received signal $r(t)$ and the modulated code $s_k(t)$. The obtained results are presented in Fig. 11. Furthermore, the main peak amplitude is shown in Fig. 12, which is obtained as the mean of the maximum peaks detected in each iteration.

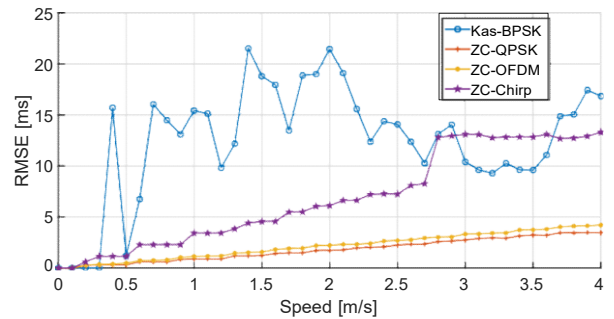

Fig. 10: RMSE detection error with Doppler effect as a function of the receiver speed.

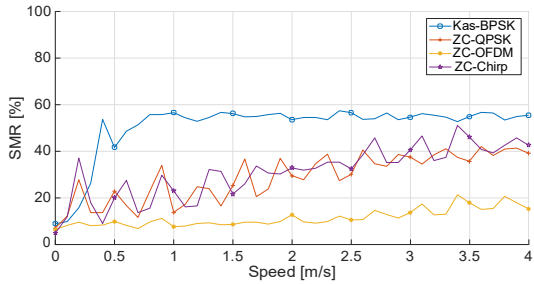

Fig. 11: Sidelobe to Mainlobe Ratio considering Doppler effect.

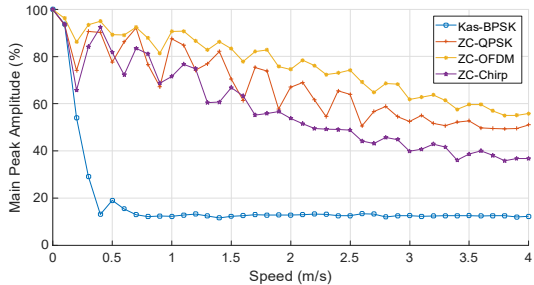

Fig. 12: Normalized main peak amplitude as a function of the receiver speed.

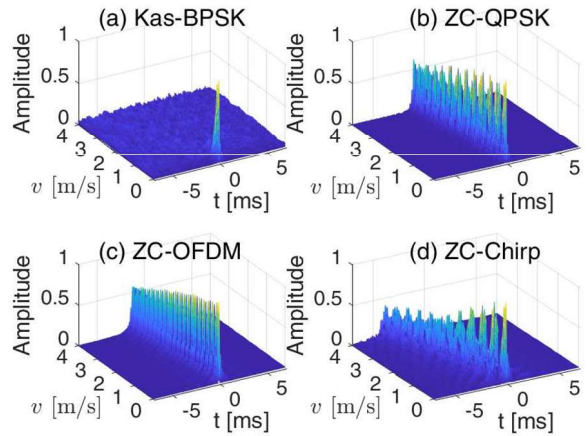

Fig. 13: Ambiguity function after Doppler effect as a function of receiver speed.

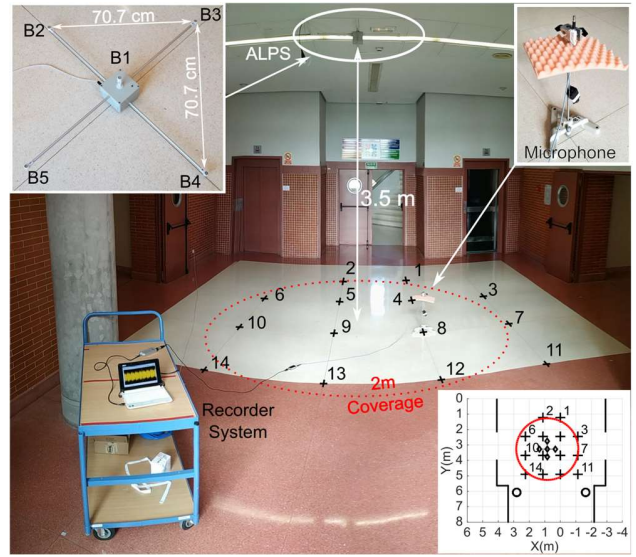

Fig. 14: Experimental setup for the static test, with the ALPS fixed at the ceiling and fourteen fixed receiver positions.

Finally, Fig. 13 shows the ambiguity function of each evaluated scheme.

## VI. EXPERIMENTAL ANALYSIS

For the real-world experiments, two different scenarios have been considered as an approach to validate the simulation results and go in depth in the performance of the addressed codes and modulation schemes. The first scenario is based on an ultrasonic ALPS placed on the ceiling and a static receiver located on the floor. It allows to test the behavior of the codes in a real ALPS and consider effects such as near-far or the performance under different ultrasonic coverage amount. The second scenario consists on a moving receiver that allows the Doppler effect evaluation. This second experiment considers both the use of a single emitter and an ALPS.

### A. Case 1: Ultrasonic ALPS considering static test positions

To carry out the static experiments, the ALPS described in [1], [34] was used. It is called LOCATE-US and consists of five hardware synchronized ultrasonic beacons (B1 to B5) that are distributed around a 70.7×70.7 cm square structure installed at a height of 3.5 m. Prowave 328ST160 [32] transducers were used for the five beacons, whose main characteristics were shown in Section IV-A. The emission is controlled by a Xilinx Zynq System-on-Chip based architecture [35], that allows the wireless configuration of the transmissions. The beacons are hardware synchronized and simultaneously transmit their corresponding signals at a 500 KHz sampling rate. At the receiver, a Brüel & Kjær 4939 Microphone [36] is used together with an Avisoft UltraSoundGate 116Hm [37] interface that allows the recording at a sampling rate of $f_s$ = 500 kHz and storage of the raw data in a computer. A Gauss-Newton hyperbolic multilateration algorithm was used to estimate the positions based on Differences in Times of Arrival (DTOA) calculation [38]. Thus, this system does not require synchronization between the emitters and the receivers.

The experimental setup is presented in Fig. 14, including a picture of the LOCATE-US ALPS. The receiver was set at fourteen different test positions, at a height of 55.1 cm above the floor. A 2 m circumference was added in red dot-line as a reference about the ultrasonic coverage. At each test point, 100 measurements were obtained per code-modulation scheme. Experiments were done under the same conditions for all evaluated cases.

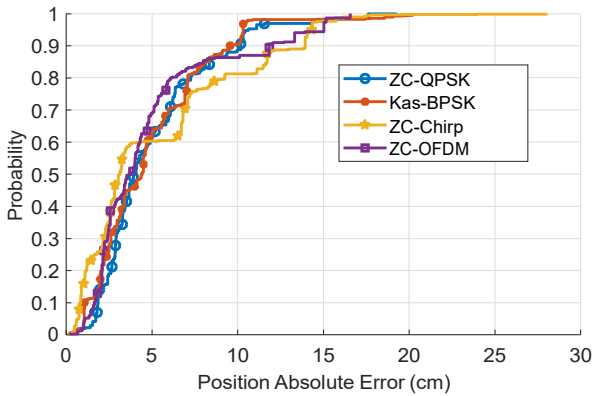

Fig. 15: CDF for the Test Points presented in Fig. 14.

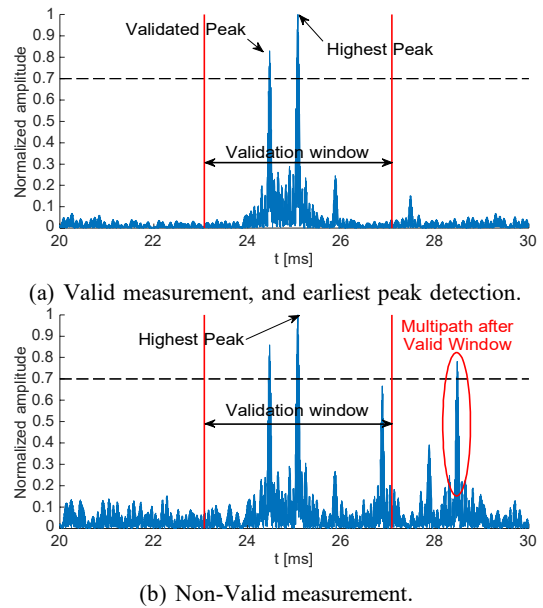

(a) Valid measurement, and earliest peak detection.

(b) Non-Valid measurement.

Fig. 16: Peak detection examples after the cross-correlation: (a) Valid measurement; (b) Non-valid measurement.

TABLE III: Percentage of valid emissions on static experiments for 100 trials per test point.

| Test Point | ZC-QPSK | Kas-BPSK | ZC-Chirp | ZC-OFDM |
|---|---|---|---|---|
| 1 | 100 | 100 | 80 | 0 |
| 2 | 100 | 100 | 100 | 5 |
| 3 | 100 | 100 | 27 | 1 |
| 4 | 100 | 100 | 95 | 100 |
| 5 | 100 | 100 | 100 | 100 |
| 6 | 100 | 100 | 100 | 100 |
| 7 | 89 | 67 | 8 | 78 |
| 8 | 100 | 100 | 100 | 100 |
| 9 | 100 | 100 | 100 | 100 |
| 10 | 100 | 100 | 100 | 19 |
| 11 | 0 | 56 | 33 | 0 |
| 12 | 100 | 100 | 8 | 0 |
| 13 | 100 | 100 | 100 | 4 |
| 14 | 100 | 0 | 100 | 0 |
| Σ | 1289 (92.07%) | 1223 (87.36%) | 1051 (75.07%) | 607 (43.36%) |

A quantification of the error for the fourteen test positions evaluated can be observed in Fig. 15, which depicts for each encoding scheme the CDF of the Euclidean distance between the ground-truth positions and the estimated ones. The CDF provides information on accuracy and precision and also allows an easy visual comparison between the different code-modulation strategies.

These results only considers those with enough information to compute a position, *i.e.* when at least the correlation peak of four of the five emitted signals are detected. At each of the 100 measurements, the correlation results from every beacon are analyzed. A peak detection algorithm has been used to discard measurements strongly affected by multipath and consider the earliest peak. The algorithm identifies the highest correlation peak and set a validation window of 4 ms around this main peak. Also, a threshold at 70% of the main peak is considered to identify further peaks. If there appear peaks out of the validation window exceeding that threshold, the measurement from the beacon is discarded. When there are no peaks outside the validation window, the algorithm looks for the first peak inside the window that exceed 70% of the main peak, and validates it as the reference for TOA or DTOA estimation. The size of the validation window and the threshold for early peak detection were experimentally obtained. Table III summarizes the percentage of valid measurements that allow to estimate a position using the Gauss-Newton algorithm.

*B. Case 2: real test with the receiver in motion*

To evaluate the performance of the proposed encoding and modulation schemes against the Doppler effect, two indoor experiments were carried out employing an electric slider [39]. The slider consists of a two meters long conveyor belt supporting a small platform where the emitter or receiver can be mounted. The transducers should be separated from the base to avoid undesired echoes. This slider is capable to reach a maximum speed of 2 m/s, with maximum acceleration values of 3 m/s$^2$. A PC software controls the position and speed of the platform according to the profile shown in Fig. 17. See [11] for further details regarding the slider configuration. In a first experiment, the LOCATE-US ALPS was used as emitter and the receiver was placed in the moving platform. The second experiment used one single emitter, that was the element in motion, while the receiver remained fixed.

*1) Case 2.1: test with the ALPS:* As mentioned before, the signal expansion or compression produced by the Doppler effect is related to the relative movement between emitter and receiver. According to (20) the effect is higher when the receiver and emitter are moving in the same direction. Thus, the ALPS was fixed in a vertical plane in such a way that the beacon one (B1) was located in front of the receiver, which was mounted over the slider platform with its movement aligned with the B1 emission direction. Fig. 18 (a) is a photo of this experimental set-up. The test starts with the receiver in front of the emitter and then moving back to reach the end of the slider.

Figs. 19(a) to (d) depict the ambiguity function for the proposed schemes, revealing their performance against Doppler

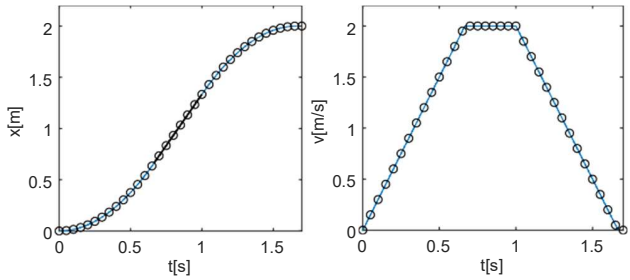

Fig. 17: Electric slider position (left) and speed (right) profiles *versus* time.

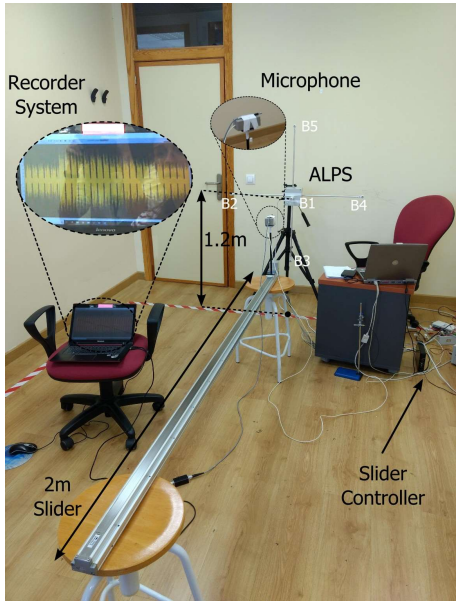

(a) Case 2.1: Experimental setup with ALPS.

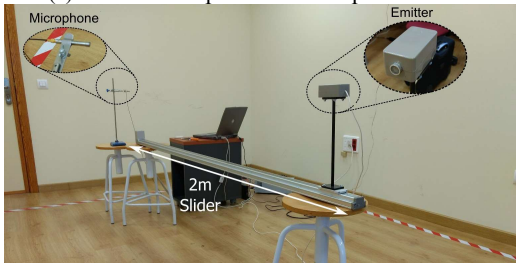

(b) Case 2.2: Experimental setup with single emitter.

Fig. 18: Experimental setup for Doppler real tests.

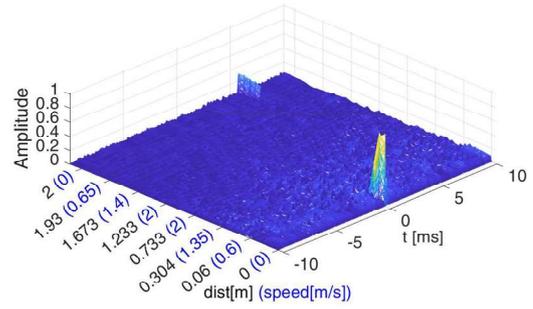

(a) Emitted signal: Kas-BPSK

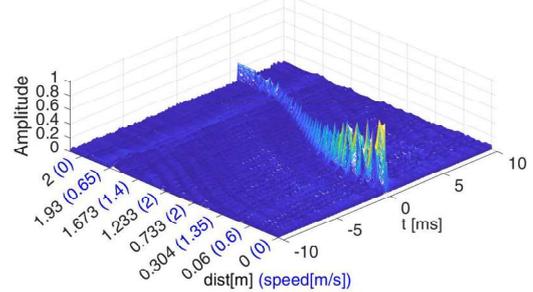

(b) Emitted signal: ZC-QPSK

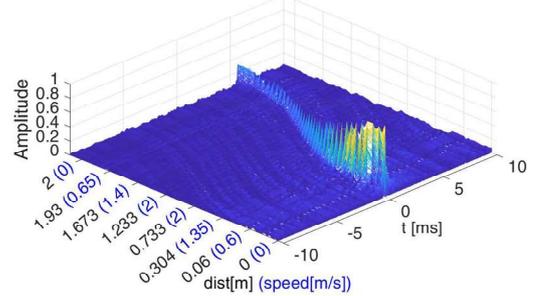

(c) Emitted signal: ZC-OFDM

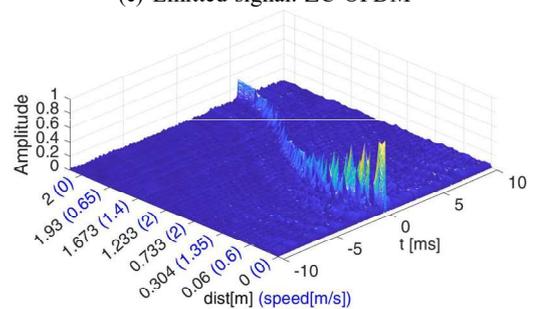

(d) Emitted signal: ZC-Chirp

Fig. 19: 3D plot correlation for beacon 1. Experiment with the receiver in motion and the ALPS as emitter.

effect and a multiple access scenario where the five beacons are simultaneously transmitting their signals. The results correspond to the correlation with the signal assigned to B1 since it is the one received with the higher quality. Even though, the other beacon correlation results are similar.

*2) Case 2.2: Single emitter:* In the second test, a single emitter is used instead of the ALPS. The emitter was mounted over the electric slider and moved forward in the direction to a G.R.A.S. 40BE [40] fixed ultrasonic microphone that was aligned straight ahead of the emitter (See Fig. 18(b)). The PC software that controls the belt movement, also controls a NI USB-6212 data acquisition card that sends the modulated signal to the emitter performing a single capture when the platform is moving at a constant speed of 2 m/s. The acquisition card stores the received data in a text file for off-line processing. Both processes are performed at a sampling frequency of 400 kS/s [11].

The aperiodic correlation is performed between the received signal and a copy of the modulated code under evaluation. The main objective is to detect the main correlation peak in spite of the emitter movement. Table IV shows the obtained results for 20 repetitions per evaluated scheme. The table includes useful information per every evaluated scheme, such as the peak width in miliseconds (the wider the peak the less resolution in the peak arrival estimation); the SMR that represents a measure of the difficulty to detect the emitted signal according to Eq. (24); and finally, the emitted energy. The energy is calculated by using Eq. (15) in two cases: considering the model of the transducer effect, and without considering it.

TABLE IV: Results with a single emitter at 2 m/s (case 2.2).

| Code / Modulation | Peak width [ms] | mean SMR | Energy w/o filter | Energy w/ filter |
|---|---|---|---|---|
| Kas-BPSK | 17.66 | 90% | 4866 | 103,04 |
| ZC-QPSK | 0.30 | 45% | 4900 | 97,31 |
| ZC-OFDM | 0.18 | 48% | 2642 | 41,12 |
| ZC-Chirp | 0.48 | 71% | 6062 | 139,19 |

## VII. DISCUSSION AND CONCLUSIONS

This research went through simulations and real test experiments to evaluate the performance of Zadoff-Chu complex sequences transmitted through phase modulation (ZC-QPSK), orthogonal frequency division multiplexing (ZC-OFDM) and chirp division multiplexing (ZC-OCDM). Besides, these modulation schemes were compared with conventional Kasami codes BPSK modulated (Kas-BPSK), often found in ALPS proposals. The presented schemes were assessed against typical effects in indoor environments.

Simulation results gathered in Fig. 5 show that Kas-BPSK, ZC-QPSK, and ZC-Chirp presents similar robustness against noise up to -15 dB, while ZC-OFDM have errors starting at about -10 dB.

Fig. 6 indicates that ZC-QPSK achieves higher robustness to near-far effect compared to the other assessed schemes, being perfectly detected for attenuations $A \gtrsim 0.4$. The ZC-OFDM scheme presents the worst performance, since in this case errors start appearing when the weakest signal is received with half the energy of the strongest one.

Fig. 8 indicates that, in a free noise environment, the selected codes achieve high immunity against multipath. ZC-OFDM is the one that exhibits a worst performance, followed by ZC-OCDM. Results for ZC-QPSK and Kasami-BPSK are quite similar, slightly better for the first one, though. The CDF increases at 0.4 ms since this is the minimum time considered for multipath arrivals.

The last simulation scenario presents the results obtained for the Doppler effect. Kasami codes are the ones that offer worst performance in terms of robustness to Doppler. Fig. 10 shows the high RMSE compared to ZC-QPSK and ZC-OFDM. This is mainly due to the significant decrease of the main peak with the speed (as can be seen in Fig. 12 and 13(a)). This decrement in the main peak amplitude implies an increase of the SMR as can be observed in Fig. 11. On the contrary, ZC-QPSK and ZC-OFDM have a main peak that can be clearly separated from the correlation sidelobes, although it suffers a small deviation that can be compensated if the receiver speed is known.

Finally, real-world experiments validate the simulated results. First, static tests were performed in an indoor environment. Results presented in Fig. 15 show that in 90% of the cases the error is lower than 14 cm, with errors below 10 cm in ZC-QPSK and Kas-BPSK cases. In addition, Table III indicates that these schemes achieve a high amount of valid emissions, around 90% of the measurements. On the contrary, ZC-Chirp and ZC-OFDM achieve about 75% and 43% of valid measurements, respectively. Also, it can be observed that positions away from the beacons offer worst results than the closer ones due to geometric dilution of precision. This performance is also in accordance to the simulation results described above. Tests with the receiver or emitter in motion have been also presented, considering two different situations. First, a multiple access with the emitter being an ALPS with five beacons, and the receiver located in a moving slider that follows a variable speed profile. Second, tests with a single emitter and receiver, at a constant speed of 2 m/s. From these experiments, the performance of ZC-QPSK and ZC-OFDM can be highlighted, since they present low degradation of the main peak in case of Doppler shifting. A linear time shift of the main peak position has been also observed, as it was in the simulated results. Table IV, which corresponds to the second experiment in motion, shows similar results than those obtained in Figs. 10 to 12, consolidating the superiority of ZC-QPSK and ZC-OFDM in terms of Doppler resilience. Kas-BPSK offers a wide peak that hinders the accurate detection of its TOA and a SMR= 90% that prevents the peak detection with speeds higher than 0.4 m/s.

The research performed indicates that ZC-QPSK is an interesting option with good performance against typical ALPS effects, including near-far, mutipath, noise and Doppler effect. Conventional BPSK modulated Kasami codes have been widely used in ALPS applications, but they achieve a limited performance in case of Doppler shifting, what constrains their use in many situations in which the receivers are in motion. ZC-OFDM is also a good option in Doppler scenarios, but is more sensitive to noise and near-far effect. ZC-OCDM offers worst immunity against impairment factors than ZC-QPSK and it also limits the multiple access due to the constrained orthogonal or cuasi-orthogonal symbols in the available transducer bandwidth.

## ACKNOWLEDGMENT

This work was supported by the Spanish Ministry of Economy and Competitiveness (MICROCEBUS project, ref. RTI2018-095168-B-C51, and SOC-PLC project, ref. TEC2015-64835-C3-2-R), Argentine National Agency for Science and Technology Promotion (PICT 2014-1875 project), and National University of Patagonia San Juan Bosco (Ref. SCyT 80020170200057UP). The authors would like to thank Dr. Chris Bleakley from UCD University (Ireland) and Dr. Jesús Ureña from UAH University (Spain) for their support and ideas at the initial stages of the work.